\pdfoutput=1

\documentclass[12pt]{article}

\usepackage{latexsym} \usepackage{graphicx} \usepackage{mathptmx}

\usepackage{amsmath} 
\usepackage{amsfonts} 
\usepackage{amssymb} 
\usepackage{amsbsy}
\usepackage{amsthm}
\usepackage{subcaption}
\usepackage{adjustbox}

\usepackage{xcolor}

\usepackage{syntax}

\usepackage{units}
\usepackage[numbers]{natbib}

\usepackage{multibib}

\usepackage[pdftex,colorlinks=true,urlcolor=blue,citecolor=black,anchorcolor=black,linkcolor=black]{hyperref}

\usepackage{hyphenat} 

\usepackage{listings}
\usepackage{color}

\definecolor{lightgray}{rgb}{.9,.9,.9}
\definecolor{darkgray}{rgb}{.4,.4,.4}
\definecolor{purple}{rgb}{0.65, 0.12, 0.82}

\lstdefinelanguage{JavaScript}{
  keywords={when, while, with, type, typeof, new, true, false, catch, function, return, null, catch, switch, var, if,
    in, while, do, else, case, break, proc, link},
  keywordstyle=\color{blue}\bfseries,
  ndkeywords={class, export, boolean, throw, implements, import, this},
  ndkeywordstyle=\color{darkgray}\bfseries,
  identifierstyle=\color{black},
  sensitive=false,
  comment=[l]{//},
  morecomment=[s]{/*}{*/},
  commentstyle=\color{purple}\ttfamily,
  stringstyle=\color{red}\ttfamily,
  morestring=[b]',
  morestring=[b]"
}

\lstdefinestyle{chai}
{
   language=JavaScript,
   extendedchars=true,
   showstringspaces=false,
   showspaces=false,
   numberstyle=\footnotesize,
   numbersep=9pt,
   tabsize=2,
   breaklines=true,
   showtabs=false,
   captionpos=b,
   numberstyle=\tiny,
   basicstyle=\footnotesize\ttfamily,
   xleftmargin=2em,
}

\lstdefinestyle{XML} {
    language=XML,
    extendedchars=true, 
    breaklines=true,
    breakatwhitespace=true,
    emph={},
    emphstyle=\color{red},
    basicstyle=\ttfamily,
    columns=fullflexible,
    commentstyle=\color{gray}\upshape,
    morestring=[b]",
    morecomment=[s]{<?}{?>},
    morecomment=[s][\color{forestgreen}]{<!--}{-->},
    keywordstyle=\color{orangered},
    stringstyle=\ttfamily\color{black}\normalfont,
    tagstyle=\color{darkblue}\bf,
    morekeywords={attribute,xmlns,version,type,release}
}

\usepackage{tgtermes}

\usepackage{fullpage}

\usepackage{authblk}

\title{A modeling and simulation language for biological cells with coupled mechanical and chemical
  processes}

\author[1]{Endre Somogyi}
\author[2]{James A. Glazier}

\affil[1]{Department of Computer Science, Indiana University}
\affil[2]{Department of Intelligent Systems Engineering, Indiana University}

\begin{document}

\maketitle

\section*{Abstract}

Biological cells are the prototypical example of active matter. Cells sense and respond to
mechanical, chemical and electrical environmental stimuli with a range of behaviors, including
dynamic changes in morphology and mechanical properties, chemical uptake and secretion, cell
differentiation, proliferation, death, and migration.

Modeling and simulation of such dynamic phenomena poses a number of computational challenges. A
modeling language describing cellular dynamics must naturally represent complex intra
and extra-cellular spatial structures and coupled mechanical, chemical and electrical processes. 

Domain experts will find a modeling language most useful when it is based on concepts, terms and
principles native to the problem domain. A compiler must then be able to generate an executable
model from this physically motivated description.  Finally, an executable model must efficiently
calculate the time evolution of such dynamic and inhomogeneous phenomena.

We present a spatial hybrid systems modeling language, compiler and mesh-free Lagrangian based
simulation engine which will enable domain experts to define models using natural, biologically
motivated constructs and to simulate time evolution of coupled cellular, mechanical and chemical
processes acting on a time varying number of cells and their environment.

\textbf{Keywords:} Biological Systems Modeling, Simulation, Spatial Hybrid Systems


\section{Introduction}
Life is the result of a complex interplay between chemical, mechanical, electrical and other
physical processes~\cite{Brodland:2015bl}. Biological cells exist in a dynamic, spatial, fluid
environment where they sense stimuli in their environment such as physical forces, chemical and
electrical signals, fluid pressure and flow, or other physical properties. Cell responses include
changes in morphology or mechanical properties, proliferation, death, differentiation, motion and
production of signals.

Mechanistic models of natural systems do not attempt to replicate physical reality in exact detail,
but describe the mechanisms which quantifiably predict specific observed behavior. Since all aspects
of a model are controlled, while many aspects of an experiment are not controllable or not
observable, simulations enable researchers to test and validate hypotheses. Simulations are often
faster than real time and are generally cheaper, safer and sometimes more ethical than real-world
experiments.

Building mechanistic models of biological phenomena can be challenging, both because of the inherent
complexity of the biological system being modeled and because models are typically written in
programming languages that do not easily describe physical concepts. Most programming languages
either represent abstractions of the underlying Von Neumann hardware architecture, or are meant to
represent models of computation (e.g., $\lambda$ calculus) rather than natural physical phenomena.

In order to enable scientists to create mechanistic models easily, we are developing a new
programming language and simulation environment, \emph{Mechanica}. Our motivating problem domain is
\emph{mesoscopic} molecular and cellular biology, where atomistic simulation approaches are not
computationally tractable and deterministic continuum approximations are not valid. Mechanica is a
physically motivated model description language and simulation engine suitable for many kinds of
mesoscopic agent-based phenomena.

Mechanica describes models in terms of objects and processes.  Objects are the physical things being
studied in our problem domain.  Objects are state-full things that exist as snapshots in time;
objects themselves are time-invariant -- they do not have any intrinsic dynamics. Processes act on
objects and change the state of objects over time. They define the dynamics and interactions of and
between objects. Objects and processes work together to define how a system evolves over
time; i.e., they describe a dynamical model.

\section{Related Work}

Languages designed for modeling dynamic physical systems differ from mainstream programming
languages in that they incorporate notions of time and/or space~\cite{Beal:2013ur} and
alleviate the need for users to think about low-level computational implementation details. Many
modeling approaches can describe different aspects of biological processes. Some modeling approaches
are well-suited for macro-scale chemical reactions, others at nano-scale atomistic dynamics, but few
approaches exist for meso-scale dynamics that allow for a natural description of coupled chemical
and mechanical processes.

Systems biology is a holistic or integrative approach that seeks to understand phenomena as a whole,
and such approaches frequently represent biological phenomena like gene regulatory, chemical or
metabolic networks. Systems biology focuses on how individual components interact and communicate
rather than investigating the internal workings of the components. Systems biology modeling
approaches tend to have constructs for representing transformation processes such as chemical
reactions and discrete events, but lack concepts to represent dynamic geometry and mechanical
processes. Systems biology models are represented with languages such as Systems Biology Markup
Language (SBML), Petri Nets or process algebras.

SBML~(\url{http://www.sbml.org/}) can describe phenomenon such as chemical reactions in a well-stirred
compartment, but has limited support for spatial processes and no concept of forces, dynamic
geometry or structural rearrangement. SBML itself has well-defined semantics for discrete events
which we adopt in Mechanica, and individual SBML models can be connected together as part of a
larger simulation environment such as discrete event simulation~\cite{Belloli:2016tu}.

Petri Nets are a modeling formalism introduced by Carl A. Petri to represent chemical
networks. Petri Nets are well-suited to describe concurrent communication and synchronization
networks. Colored Petri Nets~(CPN)~\cite{Jensen:2007ue} combine Petri Nets with a functional
programming language to create a discrete event modeling language. The integration of a functional
programming language simplifies model construction and increases the modeling capability of
CPNs. CPNs are frequently applied in systems biology, and multiple CPNs can be coupled and arranged
in a grid to simulate spatial systems such as bacterial colonies~\cite{Parvu:2015cm}.  CPNs
generally lack concepts of dynamic space and do not have a natural way to describe mechanical
processes. Certain CPN-related process algebras (like Cardelli's $3\pi$
language~\cite{Cardelli:2010wb}) can describe dynamic spatial arrangements. Mechanica also
incorporates a functional programming language similar to the way in which CPNs combine reaction
networks, discrete events and a functional programming language.

Systems biology approaches in general are not ideally suited for describing bio-mechanics, as they
do not have natural ways to represent mechanical constructs such as forces, stress, constitutive
relations, or geometry transformations.

Particle based approaches start at the nano rather than system scale and describe natural phenomena
from an atomistic perspective. Traditional MD simulators model a fixed number of spherical,
volume-excluding particles (atoms) interacting with a classical force field. MD simulators have
simple particle types and are restricted to a finite set of predefined forces. Reactive MD simulators
such as ReaxFF~\cite{2001JPCA..105.9396V}, ReaDDy~\cite{Schoneberg:2013jw}, or
SRSim~\cite{Gruenert:2010ek} extend classical MD with reacting particles. ReaxFF evolves
particles via energy or temperature constrained Newtonian dynamics, and ReaDDy and SRSim via
over-damped Langevin dynamics. Reactive MD simulators can model self-assembly of complex materials
out of particles. ReaDDy supports particle creation and annihilation, whereas particle count is
fixed in ReaxFF and SRSim. Reactive MD alters bonded relationships between particles to form
chemical complexes by by adding or removing mechanical bonds between particles. SRSim uses BioNetGen
to specify reaction rules, whereas ReaxFF and ReaDDy use their own specification format.

Reactive MD excels at describing the nano-scale, however meso- and multi-scale phenomena present
challenges. Reactive MD particles are simple, in they only have position, momentum and orientation,
and are not typically augmented with additional state variables, thus individual particles can not
model more complex things like macromolecules or cells. Model specification in Reactive MD can be
challenging because users typically specify custom forces in C++ or Java instead of a model
description language. Reactive MD does not have concepts of continuous chemical reactions or fields
like systems biology approaches.

Stochastic particle simulators such as MCell~\cite{stiles2001monte},
Smoldyn~\cite{Andrews:2010cs}, or ML-Space~\cite{Bittig:2016ep} are well-suited for spatial
biochemical simulations such as reaction-diffusion in complex geometries. These simulators move
individual rigid particles with either spatial Gillespie or Brownian dynamics as opposed to
interaction potentials as with MD. MCell is restricted to point particles with no volume exclusion,
ML-Space particles are volume excluding and Smoldyn particles can be either. Particles can react,
diffuse, be confined by surfaces and bind to membranes. Smoldyn and MCell particles are simple, but
ML-Space particles can contain other particles and support hierarchical composition.  ML-Space and
MCell models are specified with a natural and elegant rule-based approach similar to
BioNetGen~\cite{Faeder:2009jo}, and Smoldyn supports rule-based reaction network generation.
However, most stochastic particle simulators only support fixed geometries, and cannot model
mechanical or electrical processes or dynamic morphology. Multi-scale simulations are chalenging
because these simulators because do not have a way of describing continuous fields.

\hypertarget{tDPD}{Transport Dissipative Particle Dynamics} (tDPD)~\cite{Li:2015fn} combine
features from Systems Biology and MD to create a very general multi-scale modeling paradigm. tDPD
augments the simple MD particles with additional state variables such as chemical concentrations,
and adds continuous chemical reaction networks. tDPD particles can represent either solids or
fluids. tDPD is ideally suited for multi-scale models because it represents both large,
coarse-grained particles which interact with small diffusing particles represented as continuum
concentrations. tDPD is a relatively new modeling approach, hence most tDPD simulation appear to be
\emph{hard-coded} in general-purpose programming languages. To our knowledge, no model specification
language exists for tDPD. No tDPD simulation appears to support reacting particles as with Reactive
MD, nor do any tDPD simulations support discrete events as do the systems biology approaches.

Mechanica builds on the tDPD and incorporates reacting particles and discrete events, and defines a
model declaration language.

\section{Approach}\label{approach}

The \emph{Mechanica} modeling language (MML) enables construction of mechanistic models of natural
phenomena. MML is based on a \emph{functional} subset of Microsoft Typescript, and includes
conventional constructs such as variables and functions, but adopts syntactic concepts
from POVRay, a widely used spatial scene description language for geometry description, and pattern
matching rules inspired by ideas from BioNetGen, OCaml and Mathematica.

MML has two basic constructs: objects and processes, to reflect the objects and processes found in
naturally occurring phenomena. Objects in MML can be a conventional primitive type (e.g., scalars,
integers, booleans, enums), or a composite. An object is an instance of a \emph{type}. A type
defines the object's structure and layout, and a \emph{state space}, or the set of permissible
values that can be stored in the object. Objects have an identity, a \emph{symbol} which names and
refers to the object. MML introduces new types such as \emph{particles}, \emph{concentrations},
\emph{amounts}, \emph{spatial regions}, and \emph{spatial fields} with special semantics to support
spatial hybrid systems modeling.

MML represents materials as a collection of \emph{particle} and \emph{link} objects. MML particles
typically represent ``parcels'' of small pieces of real materials rather than atoms. Links are
persistent structural relationships which connect two or more material objects, and apply a
\emph{force} between the connected objects. Links generalize  bonded relationships from
molecular dynamics, with two main differences: links can be dynamically formed and broken, and their
functional form is user specified rather than fixed. This particle-based approach enables MML to
create a completely unified treatment of all materials. Solids are strongly bonded particles, fluids
are weakly bonded particles, and Mechanica can simulate any range in between.

Processes in MML are different from processes in the traditional computing sense in which processes
typically are a sequence of instructions that define a series of discrete state changes in objects.
MML processes, on the other hand, define either continuous or discrete \emph{transformations}, which
includes creation, deletion and state change. Processes map a set of inputs to a set of outputs, and
define a transformation as either a discrete atomic operation or a continuous operation. Both kinds
of processes can have \texttt{when} and \texttt{while} predicates which specify the conditions for
activating and deactivating the process. Processes have \emph{body} which calculates a probability
(real value, $[0,1]$) for discrete processes or a \emph{rate} for continuous processes. The
probability for discrete processes defines the likelihood that the process will be applied if its
predicates are true; the probability defaults to 1 if not present. The rate for continuous processes
specifies the rate of transformation of its inputs to outputs. Process and function definitions are
side effect free -- they cannot directly modify non-local symbols.

Discrete processes are similar to events in Modelica or SBML. They obey the same semantics as SBML
events, and we evaluate them using the SBML event handling algorithm we developed in
\cite{Somogyi:2014tv}. Discrete process definitions specify a \emph{pattern} of what objects to act
on, and conditions that specify when to trigger the process. Like SBML, discrete processes
instantaneously change variable values. Additionally MML discrete processes can create and destroy
object instances. (Note: this paper focuses on continuous processes and does not discuss discrete
processes in detail.)

Continuous processes are a generalization of chemical reactions and define the time evolution of
participating \emph{continuous variables} such as \emph{concentrations}, \emph{amounts}, real or
complex numbers. Continuous variables can be attached to any particle type following
\cite{Li:2015fn}. Thus, as the particles move, the attached chemical amounts are advected along with
them. Continuous processes definitions specify the \emph{rate} at which the process occurs.

Particles are the basic building blocks of materials. They have a position and optional attributes
such as spatial extent, orientation, mass. They can be point particles or have complex shapes.
Links can be applied either explicitly or implicitly. A link can be applied explicitly to a set of
material objects, and this set will remain explicitly bound until the link is removed. The link can be
removed either explicitly or by defining a \texttt{while} predicate in the link definition. Links
can also be implicitly applied to a \emph{spatial region} by specifying a \texttt{when}
predicate. Here, a process will monitor the \texttt{when} predicate, and apply the link to any set
of objects that match the predicate. In this sense, implicitly applied forces behave similarly to
long-range or non-bonded interactions in MD. 

We can define a simple particle-link system with a pair of point particles, \texttt{a} and
\texttt{b}, located at $(0,0,0)$, and $(1,0,0)$ respectively, and explicitly connect them with a
spring link as
\begin{lstlisting}[style=chai]
  a:particle(0,0,0,mass=1); b:particle(1,0,0,mass=1); 
  link(a,b){-k*(1-dist(a,b))};
\end{lstlisting}
This particular link defines a Hookean spring force, with a rest length of 1, where the \emph{dist}
function calculates the distance between two objects. Link definitions however can contain arbitrary
user-specified force definitions. This yields a very flexible model because users may specify the
interaction force to be as complex as they like. For example, a two body force could be as simple as
a Hookean interaction, or as complex as a dipole, quadrupole or octopole interaction.

A biological cell has on the order of $10^{14}$ atoms, so it is simply not computationally feasible
to model a cell at all-atom resolution. Many smaller~($< \unit[100]{nm}$) sized molecules are
soluble and readily diffuse through solution. If we are interested in micro- rather than nano-scale
phenomena, we may approximate diffusing chemicals as continuous, spatially varying
concentrations. MML has concentration and amount data types which can be attached to any material
object. Concentration values are automatically scaled by the volume of the material object that it
is attached to. Say a particle has a concentration attribute and the particle's volume increases,
then the value of the concentration decreases. As each material point moves, the attached values
move along with it. Chemical concentrations can be attached to any kind of material such as fluids
(solvents), surfaces (cell membranes) or fibers, and they can also be attached to spatial
regions. Concentration or amount types can be added to any object by adding them to the object's
definition. For example, we could create a new particle \texttt{p} with attributes \texttt{A} as a
concentration, \texttt{B} as a constant concentration (boundary value), and \texttt{C} as an amount
by:
\begin{lstlisting}[style=chai]
  p : particle { A:conc; B:const conc(5.0); C:amount(1.23); }
\end{lstlisting}
We may want to fill a region with material particles. Fluid particles have a set of forces act on
them which effectively models a fluid. To create a spherical fluid droplet, we first define a
\texttt{MaterialRegion}, which is bounded by a surface, and fill it with fluid particles. The
\texttt{fill} function determines the available space inside a region and packs it with objects of
the given type.
\begin{lstlisting}[style=chai]
MaterialRegion{
   surface:Sphere { radius:5, resolution:1 }; 
   body:fill(type=FluidParticle{radius=5});
}
\end{lstlisting}

Material regions can also have homogeneous, spatially uniform interiors if we do not specify a
body. Homogeneous regions can behave as a liquid bag if we attach a volume preservation link to the
object. When concentrations, amounts or other attributes are attached to a homogeneous material
region, their value is constant for the entire region.

\begin{figure}[hb]
  \centering
  \begin{tabular}[c]{lll}
    \hspace{-0.5cm}
    \begin{subfigure}[c]{0.35\textwidth}
      \begin{lstlisting}
type MyCell : MaterialRegion { 
   surface : Sphere { 
      radius:5, resolution:1 
   }; 
   proc (A) -> (X) {k1 * A}; 
   proc (X, 2 Y) -> (3 X) { 
      k2 * X * Y**2 
   }; 
   proc (B, X) -> (Y, D) { 
      k3 * B * X 
   };
   proc (X) -> (E) {k4 * X}; 
}
      \end{lstlisting}
    \end{subfigure}& \hspace{-1cm}
                     \begin{subfigure}[c]{0.40\textwidth}
                       \[
                         \arraycolsep=1.6pt \left[
                           \begin{array}{c}
                             \nicefrac{dA}{dt} \\
                             \nicefrac{dB}{dt} \\
                             \nicefrac{dD}{dt} \\
                             \nicefrac{dE}{dt} \\
                             \nicefrac{dX}{dt} \\
                             \nicefrac{dY}{dt} \\
                           \end{array}
                         \right] = \left[
                           \begin{array}{rrrr}
                             -1 &  0 &  0 &  0 \\
                             0 &  0 & -1 &  0 \\
                             0 &  0 &  1 &  0 \\
                             0 &  0 &  0 &  1 \\
                             1 &  1 & -1 & -1 \\
                             0 & -2 &  1 &  0  
                           \end{array} 
                         \right] \cdot \left[
                           \begin{array}{c}
                             \nu_{1} \\
                             \nu_{2} \\
                             \nu_{3} \\
                             \nu_{4} 
                           \end{array}
                         \right]
                       \]
                     \end{subfigure}&
\begin{subfigure}[c]{0.20\textwidth}
  \includegraphics[width=\textwidth]{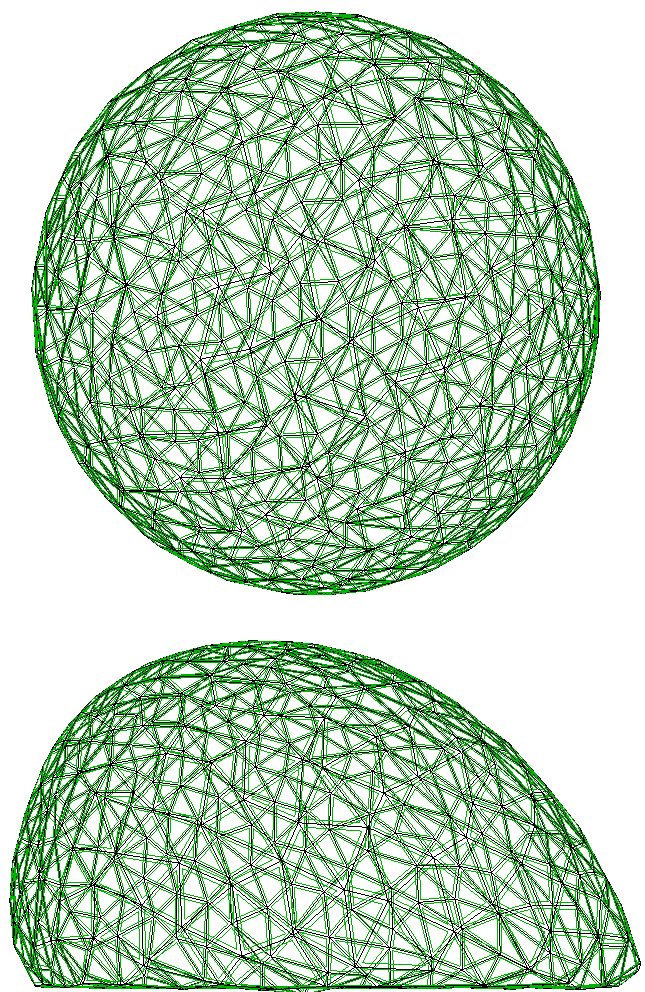}
\end{subfigure}
\end{tabular}

  \caption{Mechanica source code for creating a \texttt{MaterialRegion} derived type which is bounded
    by a spherical surface composed of a set of points and triangular faces. Four
    local continuous transformation processes define a reaction network between symbols $A, B, D, E, X$
    and $Y$. The Mechanica compiler
    generates a system of ordinary differential equations from the continuous process
    definitions. The compiler also generates a set of particles to form the bounding surface of the
    \texttt{MyCell} type. The right column shows a screenshot of the \texttt{MyCell} object in its
    initial configuration, then at a later time after it moves along a surface. 
  }\label{fig:sphere}
\end{figure}

Chemical reactions are fundamentally important in biology. Reactions define the transformation of a
set of reactants into a set of products and can be readily described by processes. A set of
continuous processes between a set of inputs (reactants) and outputs (products), such as a set of
concentrations, defines a \emph{reaction network} as in Fig.~\ref{fig:sphere}. The Mechanica compiler
reads the process definitions and generates a system of ordinary differential equations (ODEs) which
define the time evolution of reactants and products. Process definition syntax is similar to
conventional chemical reaction notation, and is based on Microsoft Typescript type
specification syntax. For example, to add a chemical reaction network to an object, one would simply write
a set of process definitions in the object definition. Process definitions begin with the
\texttt{proc} keyword, have an  optional name, a set of inputs, a set of outputs and a reaction
rate expression as in Fig.~\ref{fig:sphere}. If the object definition does not already have local
definitions corresponding to the product names from a process definition, then the process
definition implicitly adds concentration types in the object definition for these names. We do this
as a user convenience since chemical reaction networks are among the most common modeling tasks.

Processes can define intra-particle reactions, such as a flux between values located at neighboring
particles. Thus, processes can  define general reaction-transport problems. Spatial process
definitions are written and behave identically to processes confined to one object. The spatial
information is simply added in the input argument definitions. For example, to define a Fickian
(passive diffusion) flux of the value of \texttt{A} attached to material objects, \texttt{a} and
\texttt{b} of type \texttt{MyMaterial}, we would write:
\begin{lstlisting}[style=chai]
proc (a:MyMaterial.A) -> (b:MyMaterial.A) 
   when (dist(a,b) < 5) {k * (a - b)};
\end{lstlisting}

This process definition has a \texttt{when} clause that defines a cutoff distance of 5, and the
\texttt{dist} function recognizes that the values are located at different locations. 
Fickian diffusion flux of continuous valued concentrations between material points is
the same approach that \cite{Li:2015fn} used to model advection-diffusion-reaction,
$\nicefrac{dC_i}{dt} = \nabla(D_i \nabla C_i) + Q^s_i, \;\; i=1,\ldots,N$. They validated their
approach against a variety of analytical solutions and demonstrated that augmenting a particle fluid
model with scalar values at each material site incurs negligible computational overhead.

\label{sec:field}
A \emph{field} is a mapping between a value at \emph{any} spatial location and a set of values
associated with material points. In general, any continuously valued field, $A(\mathbf{r})$ can be
approximated as a sum over a set of appropriate \emph{kernel} functions,
$A(\mathbf{r}) = \sum_i f(A_i, \mathbf{r}, \mathbf{r}_i)$, where $A_i$ is some quantity located at
the source location $\mathbf{r}_i$, and $\mathbf{r}$ is the field location. This definition enables
us to define any continuous field as a function of the material. For example, the kernel function for
an electric scalar potential would be
$\nicefrac{q_i}{4 \pi \epsilon_0 |\mathbf{r} - \mathbf{r}_i|}$. In general, if the value of some
scalar quantity $A$ is known at a finite set of material points, then the value may be approximated at any
location in space $A(\mathbf{r})$ with a suitable  interpolation function~\cite{Liu:2010eo}. This
allows us to attach a chemical concentration to a material point and evaluate that concentration at
any point in space. For example, one might define a chemical compound attached to solvent points and
read the chemical concentration value at some non-solvent material point. We provide default kernels
for chemical concentrations and charge, but users are free to define their own kernels.

Most traditional languages employ symbol scoping blocks. These are regions of code that
define where a symbol (variable definition) is visible. Scoping blocks can contain other scoping
blocks.  Depending on the language, code in the child blocks can usually access variables defined
in parent blocks. Lexically scoped languages resolve non-local symbols to the scoping block in which
the function was defined. Dynamically scoped languages such as PERL resolve non-local symbols to the
scoping block in which the function is executing. 

In order to account for the spatial nature of our problem domain, we introduce a new kind of scope
resolution we call \emph{spatial scoping}. Spatial scoping is based on the idea that processes
represent a physical process, so processes always have to execute somewhere in space. The
present location of a process is implicit, much like the \texttt{this} pointer is implicit in
Java. Non-local symbols in process bodies resolve to the present spatial environment in which the
process currently runs. Spatial scoping is similar to dynamic scoping in that symbol resolution
depends on the calling context of a function; however, in spatial scoping, the calling context is a
containing spatial region.

Users typically never have to explicitly keep track of coordinate systems; instead the runtime
manages all this. All users have to know is the name of the value they want to read. Spatial scoping
treats different kinds of spatial objects uniformly. For example, consider a transformation process
which converts one substance to another, where the rate (the body of the transformation process
definition) depends on the concentration of some other substance,
\begin{lstlisting}[style=chai]
proc (A) -> (B) {k1 * A * C};
\end{lstlisting}
Here, \texttt{C} is a non-local symbol, so the runtime will search in the containing spatial
environment. This process for example could be attached to a fluid particle. This particle type may
have a \texttt{C} attribute attached to it, in which case \texttt{C} resolves to that value. If the
particle does not have such an attribute, then the compiler will check if the containing spatial
environment defines the \texttt{C} symbol. The containing spatial region may define the \texttt{C}
as a scalar, in which case the value of \texttt{C} is uniform for the entire region, and this symbol
resolves to that value. Or \texttt{C} may be defined as a spatially varying field, in which case the
compiler will generate code that reads from that field at the present location of the particle. If
this symbol cannot be resolved in any containing environment, and there is no field definition, then
the compiler will report an error.

Processes in MML are very general, being used to define not only transformations of continuous
valued objects like amounts and concentrations, but also discrete objects such as particles,
materials, and other composite objects. Processes acting on discrete objects have the identical
syntax as processes acting on continuous objects, the only difference is the semantic interpretation
of the process body. A discrete process body is expected to return a probability value, a real value
between zero and one, as opposed to a reaction rate for continuous processes. For example, line 1
defines two new types, \texttt{A} and \texttt{B}, derived from \texttt{particle}. Line 2 defines a
process which consumes two discrete particles of type \texttt{A}, and produces one particle of type
\texttt{B}. Line 3 defines a process which consumes a single object of type \texttt{B} and produces
two objects of type \texttt{A} with random probability.
\begin{lstlisting}[style=chai, numbers=left]
type A:particle{radius:1}; type B:particle{radius:2}; 
proc (a:A, b:A) -> (B) when (dist(a,b) < 5) {exp(-dist(a,b)**2)}; 
proc (a:B) -> (A, A) {rand()};
\end{lstlisting}
Processes can also be used to change the values stored in one or more attributes. Process can be
defined so that their inputs have to be in a certain state, so that they match a particular
\emph{pattern}. We borrow the record pattern matching syntax from OCaml to specify the state of an
object where a process can be applied. For example, say an object of type \texttt{A} has an
attribute named \texttt{activated} which is an enum, and we want to toggle this from
\texttt{Inactive} to \texttt{Active} when this object is near another object of type
\texttt{Activator}
\begin{lstlisting}[style=chai]
proc (a:A{activated=Inactive}, b:Activator) -> 
   (a{with activated=Active}, b) when (dist(a,b) < 5);
\end{lstlisting}
The \texttt{with} keyword is borrowed from OCaml, and specifies that the attributes in the
\texttt{with} expression should take on new values, but all other attributes should remain
unaltered. The second argument, \texttt{b} of type \texttt{Activator}, is unaffected and passes
straight through. The body of this process is omitted, which defaults the probability to 1 and tells
the compiler that this process should always be applied whenever these two particles match the
\texttt{when} pattern, i.e. are within a distance of 5.

Many proteins have \emph{binding sites} where other molecules can attach, altering the behavior of
the protein. MML introduces a new data type called \texttt{site} to represent binding sites;
any spatial object can have binding site attributes. We may for example add two binding sites to a
particle type, and use a process to bind two objects together dynamically when certain conditions
are met.
\begin{lstlisting}[style=chai]
type A:particle{s1:site(empty); s2:site(empty)};
proc (a:A{s1=empty}, b:A{s2=empty}) -> 
   (link(a.s1,b.s2){-k*(1-dist(a-b))})  when (dist(a,b) < 5);
\end{lstlisting}
This process binds the \texttt{s1} binding site of \texttt{a} to the \texttt{s2} binding site of
\texttt{b} with a new link. When a link or some other connector attaches to a binding site, the
Mechanica semantics state that that binding site should become non-empty.

Mechanica combines a \hyperlink{tDPD}{tDPD} system with discrete events and dynamic connectivity.
Any variable that is produced or consumed by a continuous process is a continuous variable; the set
of continuous variables, ${C_1, C_2, \ldots, C_n}$ form the continuous state vector
$\mathbf{C}(t)$. Each process definition body can access the present continuous variable states,
particle attributes, and constant parameters to calculate the process transformation rate.  The set
of continuous processes defines a system of ordinary differential equations (ODEs) in the form of
\begin{equation}
  \frac{d}{dt}\mathbf{C} = 
  \left[
    \begin{array}{c}
      \dot{\mathbf{C}}_{f} \\
      \dot{\mathbf{C}}_{r}
    \end{array}
  \right] =
  \left[
    \begin{array}{c}
      \mathbf{N} \cdot \nu (\mathbf{C}, \mathbf{r}, \mathbf{v}, \mathbf{p}) \\
      f(\mathbf{C}, \mathbf{r}, \mathbf{v}, \mathbf{p})
    \end{array}
  \right].
  \label{eqn:cont}
\end{equation}
The state vector comprises two partitions: $\mathbf{C}_f$ is the vector of independent continuous
variables (which participate in a transformation processes, i.e., a reaction network), and
$\mathbf{C}_r$ is a vector of variables which are defined by rate processes. The vector $\mathbf{p}$
consists of time independent parameters.  The reaction network of $m$ continuous variables and $n$
transformation processes defines the $m\times n$ stoichiometry matrix $\mathbf{N}$. Each
stoichiometric element, $\mathbf{N}_{i,j}$ is the net number of continuous variables $i$ produced or
consumed in transformation process $j$. All of the transformation process definitions combine to
form the transformation rate function $\nu$.  The second part of the state vector, $\mathbf{C}_r$, is
a set of variables which form a system of conventional ODEs, i.e. each element in this vector is
defined by a rate process. Only the Mechanica runtime can directly change variable values.

The net force on each particle is the sum of the linked, $\mathbf{F}^L$, conservative,
$\mathbf{F}^C$, random, $\mathbf{F}^R$, dissipative, $\mathbf{F}^D$, and external,
$\mathbf{F}^{ext}$ forces. The linked force is the sum of the linked relationships between the
particles. The conservative force is a soft interaction which ensures volume exclusion and keeps
materials from interpenetrating. The dissipative force represents the effects of viscosity, and the
random force represents the effects of thermal fluctuations. Integration time steps can be large
because these inter-particle forces are soft and have short-range interactions.  More details on the
conservative, random and dissipative forces can be found in \cite{Li:2015fn}. The time
evolution of particle positions, $\mathbf{r}$ and velocities, $\mathbf{v}$ is defined as
\begin{equation}
\frac{d^2\mathbf{r}_i}{dt^2} = \frac{d\mathbf{v}_i}{dt} = 
\frac{1}{m_i} \mathbf{F}_i = 
\frac{1}{m_i} \left(
\sum_{i \neq j}
\left(\mathbf{F}^L_{ij}+\mathbf{F}^C_{ij}+\mathbf{F}^D_{ij} + \mathbf{F}^R_{ij}\right) + 
\mathbf{F}^{ext}_i
\right).\label{eqn:part}
\end{equation}
The runtime calculates the time evolution of the model via \emph{time slicing} -- time is
partitioned into a series of discrete time steps.  Each time step has three phases: (A) evaluate and
apply the \texttt{when} and \texttt{while} predicates of the discrete processes determine if they
should be triggered, and apply them. Links are attached and detached in phase 1 according to their
\texttt{when} and \texttt{while} predicates. (B) integrate the continuous state vector variables
according to eqn.~\ref{eqn:cont} and (C) integrate the particle positions according to
eqn.~\ref{eqn:part}.

\subsection{Mechanica model of chemotaxis}
Chemotaxis is a biological process which describes the motion of a cell towards or away from a
chemical signal. We use chemotaxis as a biological example to illustrate the coupling of chemical
and mechanical processes and demonstrate how we can represent a few chemotaxis mechanisms in
Mechanica. Cells undergoing chemotaxis can sense spatial gradients in chemoattractant
concentrations. Chemoattractants are frequently secreted from some source, such as a bacterium at a
specific location. Here, we define a rule to secrete a chemoattractant from a fixed point source in
space. We first define the the model environment with an an explicit solvent. We write a rule in the
top level spatial region which fills it with explicit solvent particles (line 1). This rule states
that any space that is not filled with sub-objects gets filled with solvent particles. This rule
creates a symbol called \texttt{solvent} which is of type water. \emph{CXC} is a common biological
chemoattractant, which we will secrete from a point source located at $(20,20,0)$ (line 2). This
rule increases the concentration of CXC on the solvent particles whenever they are within a distance
of 5 from the point $[20,20,0]$. Because the CXC symbol is listed as a product, the rule implicitly
adds the symbol \texttt{CXC} as a chemical concentration attribute attached to the solvent
particles.  The solvent will transport the CXC along with it, but we want to allow the CXC to
diffuse through the solvent. We can use the built-in diffusion rule or we can write our own. Say we
want simple Fickian diffusion between solvent particles (line 3). This process defines a
transformation process to represent diffusion of solute between neighboring particles. It has a
flux rate of $k * (a - b)$, where $k$ is a constant, and $a,b$ are concentrations of CXC located at
neighboring particles.
\begin{lstlisting}[style=chai, numbers=left]
solvent:fill(type=Water); 
proc () -> (b:solvent.CXC) when (dist([20,20,0], b) < 5) 
   {aRateConstant}; 
proc (a:solvent.CXC) -> (b:solvent.CXC) 
   when (dist(a,b) < 5) {k * (a - b)};
\end{lstlisting}
Our cell model will only define a single basic cell object that sticks to a horizontal surface. In
order to represent a single cell, we can define a basic \texttt{MaterialRegion} derived type as in
Fig.~\ref{fig:sphere}, where the initial cell is shown in the top-right. We can place a single
instance centered at $[0,0,2.5]$ via
\begin{lstlisting}[style=chai]
mycell:MyCell{origin:[0,0,2.5]};
\end{lstlisting}
Cells frequently stick to the surface that they're on. We can represent this adhesion with a link
that causes the cell's surface to stick to the base surface.  Stickiness tends to be asymmetric, so
we can write the link rule to attach at a shorter distance than when it releases. We can think of it
as similar to a Velcro adhesion, where the two components must be close to activate, but the force
can hold on a bit longer when its stretched. In addition to the \texttt{when} clause, links can also
have a \texttt{while} clause which can define a deactivation condition that differs from the
activation condition. This rule creates links which connect the particles of the cell's surface to
the floor of our simulation domain.
\begin{lstlisting}[style=chai]
link(a:mycell.surface,b:BoundingPlanes[FLOOR]) 
  when (dist(a,b) < .5) while (dist(a,b) < 2) 
  {-k*(dist(a,b))}
\end{lstlisting}
In our simple model we approximate the cytoplasm as water. We fill the body of the cell with
water particles by adding the following line to the cell definition.
\begin{lstlisting}[style=chai]
mycell.body:fill(type=Water); 
\end{lstlisting}
A biological cell's surface often contains many chemoattractant receptors.  Essentially these bind
and degrade diffusing chemoattractant molecules in the fluid medium around the cell, and initiate a
spatially varying signaling cascade in the cell.  We observe that the receptor can be either active
or inactive, and only active receptors initiate signaling inside the cell. We can model these
observations with the following rules: Line 1 creates a surface bound concentration of active
receptors \texttt{mycell.body.ARectpt}, with an initial concentration of 0.0. Line 2 represents
passive diffusion of solvent CXC to cell's surface, with a diffusion rate of $k * (a - b)$. Lines
3-4 define a reaction that consumes CXC on the surface and produces active receptor
\texttt{ARecpt}. There are a finite number of receptors, so we want this reaction to saturate, hence
the form of the reaction rate.
\begin{lstlisting}[style=chai, numbers=left]
mycell.surface.ARectpt:conc(0.0);
proc (a:solvent.CXC) -> (b:mycell.surface.CXC) 
   when (dist(a,b) < 5) {k * (a - b)};
proc (a:mycell.surface.CXC) -> (b:mycell.surface.Recept) {
   k2 * a * exp(-b**2);
};
\end{lstlisting}
Signaling cascades inside the cell usually involve many stages, but here we assume that an internal
compound Rho (known to induce actin polymerization) can be activated by an active surface receptor,
\texttt{ARecept}. Many chemotaxis models propose that actin filament polymerization in the leading
edge of a migrating cell is principally responsible for lamelipodium extension. We approximate this
mechanism with a set of spring-like forces that push the cell nucleus away from the cell surface.
In the presence of active Rho, the springs elongate and adjust their applied force between the
nucleus and the surface particles; otherwise they have no effect. We model Rho as a cytosol
diffusing compound which decays over time.  Line 1 creates a new concentration called \texttt{Rho}
attached to the body of the cell (the cytoplasm).  Line 2 produces \texttt{Rho} on cytoplasm
particles at a rate proportional to surface \texttt{ARecpt} concentration. This is an example of
spatial scoping, where the reaction occurs at a cytoplasm fluid particle (the reaction is producing
\texttt{Rho} on the cytoplasm particles), so symbols referring to concentrations \emph{not} attached
to the particle resolve to the spatial field produced by these concentrations. The
\texttt{mycell.surface.ARecpt} symbol resolves to a value that approaches the exact surface
concentration value when the cytoplasm particle is near, but falls off rapidly if the cytoplasm
particle's distance away from the surface is high.  Line 3 defines a passive diffusion of
\texttt{Rho} between neighboring cytoplasm fluid particles.  Line 4 cause Rho to decay at a rate of
$k \cdot [Rho]$, or proportional to the concentration of Rho.  Lines 5-7 define a set of links that
attaches the cell nucleus particle to the cell surface particles.  Spatial scoping evaluates
non-local symbols in link definitions at the midpoint of the link.  Thus, when cytoplasm carrying
high concentrations of \texttt{Rho} move near the link midpoint location, the \texttt{Rho} symbols
evaluates to a high value, which causes the spring force to increase. If the concentration of Rho at
the link location decreases, say, the link moves away or the Rho decays at this location, the spring
force will contract, causing the cell membrane to retract.
\begin{lstlisting}[style=chai, numbers=left]
mycell.body.Rho:conc(1.5);
proc () -> (a:mycell.body.Rho) {k * mycell.surface.ARecpt}; 
proc (a:mycell.body.Rho) -> (b:mycell.body.Rho) 
   when (dist(a,b) < 5) {k * (a - b)};
proc (a:Rho) -> () {k * a}; 
link(a:mycell.nucleus, b:mycell.surface) {
  -k * (dist(a,b) - restLength + k2 * Rho);
}
\end{lstlisting}
\subsection{Simulator Details}\label{compiler-details}
The Mechanica simulator is loosely based Model-View-Controller paradigm. It is comprised of five
key objects: Simulator, Model, View, Integrator, and Compiler. The Simulator is the top-level object
which orchestrates the interaction of the others. It uses the Compiler to read a source MML
document(s) and generate a Model. The Model contains the  compiled form of the MML model  and the
present state. The Integrator calculates the time-evolution of the Model. The View
displays the present model state in an OpenGL window. The Model notifies the view when its state
changes so that the View can update the display.

The front end of the Mechanica compiler is a conventional recursive descent parser which generates
an abstract syntax tree (AST). The semantic analysis phase of the compiler, however, differs from
procedural programming language compilers. Unlike procedural languages where there is a reasonably
close mapping between language statements and the resulting machine instructions, MML statements
specify rules and relationships rather than explicit instructions. The Mechanica compiler analyzes
the model processes and rules and use this information to generate the two time evolution systems of
equations, eqn.~(\ref{eqn:cont}) and eqn.~(\ref{eqn:part}).  The Mechanica compiler performs
semantic analysis of the continuous process definitions using techniques we established in
\cite{Somogyi:2014tv}. The Mechanica compiler presently generates C code, and uses GCC to compile
this into a shared library. The compiler then creates a new Model object which loads this shared
library. Although MML object definitions appear similar to a C struct, our in-memory
storage is different. All the continuous variables are stored in a single contiguous array, and the
particle position and velocity vectors are also stored in contiguous arrays.

The Mechanica integrator uses CVODE~\cite{hindmarsh2005sundials} to calculate the time
evolution of the continuous variables. We chose the contiguous memory layout to match CVODE. At each
time step, CVODE operates in-place on this shared memory block. We also use CVODE's root finder to
check if any \texttt{when}/\texttt{while} predicates should trigger  discrete
processes. The Mechanica Integrator uses a modified version of Pedro Gonnet's mdcore
library~(\url{http://mdcore.sourceforge.net/}) to calculate the time evolution of the particle
positions. The contiguous array storage of the particles and velocities also matches mdcore's layout,
thus mdcore operates in-place with no memory copying. mdcore uses the compiled MML model force
function to calculate inter-particle potentials. The integrator uses the distance cut-off
information from the link predicates to specify a potential cut-off distance for mdcore, which
mdcore then uses to construct optimal Verlet and cell lists.  Unlike traditional molecular dynamics
or particle systems, the particle count in Mechanica models changes over time. The runtime initially
allocates a large block, more than the initial particle count, and as particle count increases, it
performs a realloc on that block. We are investigating more efficient approaches for dealing with
this challenge.

\section{Conclusion}
We are developing the first known modeling language which can represent generalized continuous
transformation processes, forces, fluid flow, and material state transition in the same
language. The Mechanica language enables users to create physically motivated models of complex
natural phenomena using mechanistic instead of computational building blocks. Mechanica is currently
under active development, all source code and binaries \emph{will be made} available on the
Mechanica website, (\url{http://www.mechanica.org}) under a permissive open source license (BSD/GPL
dual license).

\section*{acknowledgments}
We acknowledge generous financial support the National Institutes of Health, National Institute of
General Medical Sciences, National Institute of Environmental Health Sciences and National Institute
of Biomedical Imaging and Bioengineering, grants U01 GM111243, R01 GM076692 and R01 GM077138. We
thank Dr. Marie Gingras, Dr. Amr Sabry and Dr. Paul Macklin for their discussion and insights.


\end{document}